\documentstyle[aaspp4]{article}
\begin{document}
\title{ A Note on TeV Cerenkov Events as Bose-Einstein Gamma Condensations}
\author{Amir Levinson\\ School of Physics and Astronomy, Tel Aviv University,
Tel Aviv 69978, Israel}
\begin{abstract}
The idea that the TeV air showers, thought to be produced by $>10$ TeV 
gamma rays from Mrk 501, can be mimicked by coherent bunches of 
sub-TeV photons is reexamined, focusing on fundamental considerations.
In particular, it is shown that the minimum spot size of the beam of pulsed 
TeV photons arriving at Earth is on the order of a few kilometers, unless a 
lens with certain characteristics is placed between the TeV laser and Earth.  
The viability of the laser production mechanism proposed by 
Harwit et al. (2000) is also reassessed.
\end{abstract}
          

It has been argued recently (e.g., Coppi \& Aharonian 1999) 
that the detection of TeV photons from
Mrk 501 at energies well above 10 TeV (Hayashida, et al. 1998; 
Aharonian et al. 1999) places severe constraints
on the diffuse extragalactic IR background, and may be particularly
problematic (Protheroe \& Meyer, 2000) in view of recent determinations of 
the IR background by
various experiments.  Motivated by this consideration, Harwit et al. (1999)
proposed that the observed air showers, that are commonly interpreted
as due to single $>10$ TeV gamma rays, can be produced by coherent 
bunches of sub-TeV photons that are not absorbed by pair production on
the extragalactic IR background.  What they envisaged essentially is that some 
mechanism in the source (Mrk 501) produces pulsed, sub-TeV lasers that, 
when interacting
with the Earth's atmosphere could mimic TeV air shower events.  This 
hypothesis has been tested subsequently by the HEGRA collaboration 
(Aharonian et al. 2000), who claimed that it can be rejected on the basis
of a comparison of the energy dependent penetration depth in Earth's 
atmosphere of TeV photons from Mrk 501 with the penetration depth of
photons from the Crab Nebula.  In view of the growing interest in this
scenario it is worth reexamining its viability.
Below, we discuss the fundamental limitations of such a TeV laser. 

What are the basic requirements from the system under consideration? 

i) A typical Cerenkov flash produced by a TeV air shower lasts for about
several ns.  This implies that the width of the TeV pulse produced by the 
laser, $\Delta t$,  should not exceed this timescale, and that 
temporal coherence must be maintained over a time $>\Delta t$.  The 
corresponding light crossing time, $c \Delta t$, is of the order of 
several meters.  In principle, however, the dimension of the system should
not be restricted to this scale.  In laboratory lasers, for instance, 
pulse durations as short as the decay time of the lasing substrate 
(which can be shorter by many orders of magnitude than the light crossing 
time of the cavity, though typically larger than the beam diameter) 
can be achieved using e.g., mode locking or Q switching 
methods (which require modulation of either the pumping 
rate or the refraction index in the cavity) 
(e.g., Svelto 1998).  Although it is difficult
to envisage how this situation can be accomplished under astrophysical 
conditions, the requirement that the size of the system would not 
exceed the pulse width does not seem to be fundamental.  
Moreover, if the laser mechanism involves relativistic motion the pulse 
can be further compressed, owing to time dilation effects.

ii) The spot size of the TeV laser beam illuminating Earth should be within
the angular resolution of current TeV experiments, otherwise the shower 
image will differ from that expected to be produced by a single TeV 
photon. (If the spot is resolved, it can give rise to a shower image that 
may resemble that of a cosmic ray shower.  Such an event is likely to be 
rejected.)  For a typical angular resolution of 0.1$^\circ$, and shower 
hight of say 10 km, this yields a spot size $<20$ meters.  As shown below 
this requirement places a stringent constraint on the system.

Harwit et al. argued that the arriving bunched photons would spread
over a distance $\Delta y\simeq h/\Delta p_y \simeq \lambda L/D$, where 
$\Delta p_y$ is the uncertainty in transverse momentum, which equals 
approximately the product of the photon momentum, $h/\lambda$, and the 
angular size of the source subtends at Earth, $D/L$, with $D$ being
the characteristic source dimension and $L$ the distance between Earth
and the source (Mrk 501).
Taking $D$ to be on the order of a Schwarzschild radius of a supermassive 
black hole yields $\Delta y$ in the range between $10^{-1}$ to $10^{-4}$ cm
for the expected range of black hole masses, which, as they argued, is 
negligible in Cerenkov detection.  It should be noted, however,
that $\Delta y$ is essentially the size of the diffraction wing of a beam
having a diameter $D$ at a distance $L$ from the beam waist (see Fig. 1) 
and not the spot size, which for the above parameters is practically 
$D$. Thus, if the dimension of the photon bunch (or Bose Einstein condensate,
as termed by those authors) is indeed $D\sim r_g$, then in the absence of an 
additional optical system (see below), the arriving photon bunch would 
spread over a similar scale, not $\lambda L/r_g$.  We suspect though that 
the picture envisioned by Harwit et al. is that each condensate is well 
localized, so that the spatial dimension of each bunch at the source 
is not $D$ but very 
small.  In that case the uncertainty in transverse momentum of a photon 
in the bunch is related to the latter scale and is much larger than
that quoted above.  The question then is, what is the minimum 
spot size that can be achieved given the distance between Earth and 
Mrk 501?  

\begin{figure}
\centerline{\epsfxsize=3.3in\epsfbox{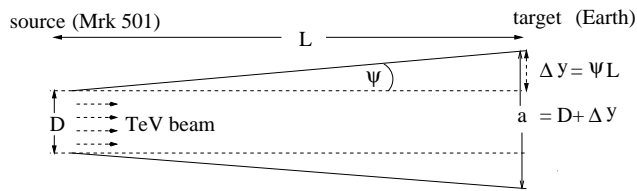}}       
\caption{Schematic illustration of the propagation of a coherent 
TeV beam having a diameter $D$ at its waist.  The diffraction angle 
of the beam is $\psi=\lambda/D$, where $\lambda$ is the corresponding
wavelength, and the beam spot size a distance $L$ away is $a = D+\psi L$.}
\end{figure}

Consider some apparatus that produces a pulsed TeV laser beam having 
a diameter $D$ at the beam waist (see 
Fig. 1). The diffraction angle of the beam is $\psi=\lambda/D$, where 
$\lambda =2\times 10^{-17} (\epsilon/1 {\rm TeV})^{-1}$ cm is the wavelength 
of the laser at its spectral peak, and $\epsilon$ is the corresponding 
energy.  At a distance $L$ from the laser the beam spot size, $a$, is the 
sum of the waist spot size and the size of the diffraction wing: 
\begin{equation}
a=D+\psi L=D+(\lambda/D)L.
\end{equation}
For a target at a fixed distance $L$ from the laser, the minimum beam spot 
size $a_{min}$ can be obtained by minimizing $a$ with respect to $D$; 
that is taking $da/dD=0$.  This yields $D=\sqrt{\lambda L}$ and 
\begin{equation}
a_{min}=2\sqrt{\lambda L}\simeq 10^5 (L/100 {\rm Mpc})
(\epsilon/1{\rm TeV})^{-1}\ \ \ {\rm cm}.
\end{equation}
Taking $L$ to be the distance from Earth to Mrk 501 ($L=130$ Mpc assuming 
$h_o=0.65$), we conclude that the arriving pulse of sub 
TeV photons would spread over a distance of at least several kilometers.

The spot size can in principle be reduced if a lens is placed between the 
source and Earth.  In the optimal case the radius of the lens should be
comparable to the beam radius, and its focal plan should intersect
Earth.  Unser such conditions the size of the spot illuminated by 
the TeV laser 
is diffraction limited (assuming complete coherence).  Denoting 
by $D_l$ the lens diameter and 
by $L_l$ its distance from Earth, and requiring a spot size smaller 
than say 10 meters, yields a minimum lens diameter of 
\begin{equation}
D_{lmin}=10^7 (L_l/100 {\rm Mpc})(\epsilon/1{\rm TeV})^{-1}\ \ \ {\rm cm}.    
\end{equation}
This is larger than the gravitational radius of a stellar mass object.
As an illustrative example consider lensing of the TeV beam by a point 
mass located on the axis of the beam a distance $L_l$ from Earth.  In
that case only rays for which the impact parameter lies in the range 
between $b_{+}$ and $b_{-}$, where $b_{\pm}\simeq 2\sqrt{r_g L_l}\pm d$,
$d$ being the diameter of the beam spot on Earth, and $r_g$ is the 
gravitational radius of the lens, will be deflected by the required 
angle.  For typical parameters, this constitutes only a tiny fraction 
of the beam photons and, therefore, a point mass cannot provide the 
required lens.  What seems to be needed is some extended object with
a density profile such that the refraction index of the lens would be
independent of the lens radius.  Perhaps high velocity clouds?

We conclude by briefly commenting on the TeV laser production mechanism
discussed by Harwit et al. 
These authors suggested that inverse Compton scattering of OH or H$_2$O
megamaser photons by a relativistic beam of nonthermal (in the comoving 
frame) electrons, may provide a mean for producing a coherent TeV pulse.  
We stress
that spontaneous scattering is a process that tends to destroy coherence;
a single seed photon has a probability of being scattered into many different
directions with different phases.  (It is worth noting that the
occupation number of scattered photons will be reduced not only by the 
ratio of bandwidths, as pointed out by Harwit et al., but also by the ratio
of angular sizes of the beam of scattered photons and the megamaser beam, 
which can in principle be large.)  The fact that the occupation number 
may largely exceed unity does not imply that the emission is coherent,
unless the phases of emitted photons are somehow correlated.  In order to 
mimic TeV air showers, it is required that several photons will be 
scattered into a solid angle, henceforth denoted by $\Delta \Omega$, 
not larger than that subtends at the source by the maximum spot size 
allowed in the Earth's atmosphere, and will have a time spread of no more
than several ns.  Since the scattering process is unsynchronized, it implies 
that the characteristic dimension of the emitting volume should not 
exceed $\Gamma^2$c$\Delta t$, where $\Delta t\sim 10$ ns is again the 
time scale of a typical Cerenkov shower,
and $\Gamma$ is the bulk Lorentz factor of the electron beam.
Using the number density of maser photons adopted by Harwit etal., 
$n_{ph}\sim10^{10}$ cm$^{-3}$, yields a scattering rate per electron of 
$\nu_{sct}=\sigma_T cn_{ph} \sim 2\times10^{-4}$ s$^{-1}$.  The total 
scattering rate by a beam section of volume $\Delta V\sim 
(\Gamma^2 c\Delta t)^3$ containing relativistic electrons having 
density $n_e$ is then $\nu_{sct}\Delta V n_e
\sim 2\times 10^{16}$ ($\Delta t$/$10$ ns)$^{3}$($\Gamma/100$)$^6$
($n_e$/$10^2$ cm$^{-3}$) 
s$^{-1}$.  Now, only a fraction $\Delta \Omega/\Delta \Omega_b$ will be
scattered into a solid angle $\Delta\Omega$, where $\Delta \Omega_b$ denotes
the solid angle of the TeV photon beam, which in the case of scattering by 
an electron beam having bulk Lorentz factor $\Gamma$ is on the order of 
$\Delta\Omega_b\sim \pi\Gamma^{-2}$. For a maximum spot size $d$ we thus
obtain: $\Delta \Omega/\Delta \Omega_b\simeq 10^{-43}$ ($\Gamma/100$)$^2$(
$d/10^3$ cm)$^2$($L/100$ Mpc)$^{-2}$.
Consequently, the rate at which the
maser photons in the volume $\Delta V$ are scattered into a given direction
within a solid angle $\Delta\Omega$ is approximately $10^{-27}$ s$^{-1}$,
about 35 orders of magnitudes smaller than what required.

I thank M. Segev, D. Eichler, D. Maoz, and A. Gal Yam for enlightening 
discussions.
Support from the Israel Science Foundation is greatly acknowledged.


\end{document}